\documentclass[aps,prd,nofootinbib,showkeys]{revtex4}

\begin{document}

%\markboth{S.~Dhasmana and Z.~K.~Silagadze}
%{Finsler space-time in light of Segal's principle}

%%%%%%%%%%%%%%%%%%%%% Publisher's Area please ignore %%%%%%%%%%%%%%
%\catchline{}{}{}{}{}
%%%%%%%%%%%%%%%%%%%%%%%%%%%%%%%%%%%%%%%%%%%%%%%%%%%%%%%%%%%%%%%%%%%

%\title{\bf FINSLER SPACE-TIME IN LIGHT OF SEGAL'S PRINCIPLE}
\title{Finsler space-time in light of Segal's principle}

%\author{\footnotesize S.~DHASMANA}
%\address{Novosibirsk State University, Novosibirsk 630 090, Russia \\
%dx.shailesh@gmail.com}

%\author{\footnotesize Z.~K.~SILAGADZE}

%\address{Budker Institute of Nuclear Physics and Novosibirsk State University 
%\\ Novosibirsk 630 090, Russia \\ Z.K.Silagadze@inp.nsk.su}

%\maketitle

%\pub{Received (Day Month Year)}{Revised (Day Month Year)}

\author{S.~Dhasmana}
\email{dx.shailesh@gmail.com}
\affiliation{Novosibirsk State University, Novosibirsk 630 090, Russia }

\author{Z.K. Silagadze}
\email{Z.K.Silagadze@inp.nsk.su}
\affiliation{ Budker Institute of
Nuclear Physics and Novosibirsk State University, Novosibirsk 630
090, Russia }

\begin{abstract}
ISIM(2) symmetry group of Cohen and Glashow's very special relativity is
unstable with respect to small deformations of its underlying algebraic 
structure and according to Segal's principle cannot be a true symmetry
of nature. However, like special relativity, which is a very good description
of nature thanks to the smallness of the cosmological constant, which 
characterizes the deformation of the Poincar\'{e} group, the very special 
relativity can also be a very good approximation  thanks to the smallness of 
the  dimensionless parameter characterizing the deformation of ISIM(2).
\end{abstract}

\maketitle

\section{Introduction}
Einstein's two famous postulates imply the special theory of relativity 
and Minkowski metric of space-time only when it is assumed that space is 
isotropic. If we abandon this assumption, then a Finsler generalization of the 
special theory of relativity with the corresponding Finsler metric naturally 
arises, as will be shown below.

For historical reasons, this Finsler metric will be called the 
Lalan-Alway-Bogoslovsky metric. It is characterized by one extra 
dimensionless parameter $b$. When $b$ tends to zero, in general, we do not 
obtain the special theory of relativity  as a limiting case, since a 
preferred light-like direction $n^\mu$ can survive this limit and the resulting
space-time symmetry  group will be not the Poincar\'{e} group but its 
subgroup ISIM(2). Cohen and Glashow suggested \cite{1} that it is possible 
that just this particular ISIM(2), and not the Poincar\'{e} group, is a true 
space-time symmetry group. The corresponding theory is called "very special 
relativity" (in the bibliography we provide an incomplete list of modern 
references on the subject \cite{1A}). 

Drawing an analogy with the cosmological constant, it can be argued that in 
reality $b$ is not zero, but very small. In this case, most likely, it will be 
impossible to detect the Finslerian nature of space-time in laboratory 
experiments.  Nevertheless, the question of the true value of the parameter 
$b$ is as fundamental as the question of why the cosmological constant 
$\Lambda$ is so small. Perhaps both questions are just different parts of the 
same puzzle.

\section{Anisotropic special relativity}
Both Einstein \cite{2} and Poincar\'{e} \cite{3,4} obtained $\lambda$-Lorentz 
transformations
\begin{equation} 
x^\prime=\lambda(V)\gamma\left(x-V\,t\right ), \;\;
y^\prime=\lambda(V)y,\;\;
z^\prime=\lambda(V)z,\;\;
t^\prime=\lambda(V)\gamma\,\left (t-\frac{V}{c^2}\,x\right ),
\label{eq1}
\end{equation}
and then both argue that $\lambda(V)=1$. The arguments of Einstein were  
physical, while Poincar\'{e}'s reasoning was more formal and was based on the 
analysis of the full Lorentz's group including not only boosts, but also 
spatial rotations. In particular, Poincar\'{e} notes that the group property
of transformations (\ref{eq1}) (which is equivalent to the relativity 
principle) implies the following multiplication law (which first appeared in 
Poincar\'{e}'s May 1905 letter to Lorentz \cite{5}):
\begin{equation}
\lambda(V_1\oplus V_2)=\lambda(V_1)\lambda(V_2),
\label{eq2}
\end{equation}
Note that (\ref{eq1}) and (\ref{eq2}) determine the most general form of 
the boost along the $x$-direction, compatible with Einstein's two postulates. 
The solution of the functional equation (\ref{eq2}) is hindered by the 
complexity of the relativistic velocity addition law. However the natural 
parameter for special Lorentz transformations is not the velocity $V$, but the 
rapidity $\psi$ \cite{6}, defined by
\begin{equation}
\tanh{\psi}=\beta=\frac{V}{c}.
\label{eq3}
\end{equation}
For special Lorentz transformations, rapidities are additive, and the 
functional equation (\ref{eq2}) takes the form of the Cauchy exponential 
functional equation
\begin{equation}
\lambda(\psi_1+\psi_2)=\lambda(\psi_1)\,\lambda(\psi_2).
\label{eq4}
\end{equation}
It is a well-known fact \cite{7} that all continuous solutions of (\ref{eq4})
have the form  
\begin{equation}
\lambda(\psi)=e^{-b\psi}=\left(\frac{1-\beta}{1+\beta}\right )^{b/2},
\label{eq5}
\end{equation}
where $b$ is  some real constant. As a result, $\lambda$-Lorentz 
transformations (\ref{eq2}) take the form
\begin{eqnarray} && 
x^\prime=\left(\frac{1-\beta}{1+\beta}\right )^{b/2}
\gamma\left(x-V\,t\right ), \;\;
y^\prime=\left(\frac{1-\beta}{1+\beta}\right )^{b/2}y,\nonumber \\ &&
z^\prime=\left(\frac{1-\beta}{1+\beta}\right )^{b/2}z,\;\;
t^\prime=\left(\frac{1-\beta}{1+\beta}\right )^{b/2}
\gamma\,\left (t-\frac{V}{c^2}\,x\right ).
\label{eq6}
\end{eqnarray}
The special isotropy requires $b=0$ and this is the situation corresponding
to the usual special relativity.

To find a generalization of the relativistic interval, which remains invariant
under transformations (\ref{eq6}), it is convenient to introduce light-cone 
coordinates \cite{8}
\begin{equation}
u=ct+x,\;\;\;v=ct-x,
\label{eq7}
\end{equation}
which transform as follows
\begin{equation}
u^\prime=e^{-(1+b)\psi}\,u,\;\;\;v^\prime=e^{(1-b)\psi}\,v,\;\;\;
y^\prime=e^{-b\psi}\,y,\;\;\;z^\prime=e^{-b\psi}\,z.
\label{eq8}
\end{equation}
Using (\ref{eq8}), we can easily find that
\begin{eqnarray} &&
\left(\frac{v^\prime}{u^\prime}\right )^b u^\prime v^\prime=
\left(\frac{v}{u}\right )^b u v,\;\;
\left(\frac{v^\prime}{u^\prime}\right )^b y^{\,\prime\,2}=
\left(\frac{v}{u}\right )^b y^2,\nonumber \\ &&
\left(\frac{v^\prime}{u^\prime}\right )^b z^{\,\prime\,2}=
\left(\frac{v}{u}\right )^b z^2,\;\;\frac{u^\prime v^\prime}{y^{\,\prime\,2}}
=\frac{u v}{y^{2}},\;\;\frac{u^\prime v^\prime}{z^{\,\prime\,2}}
=\frac{u v}{z^{2}}.
\label{eq9}
\end{eqnarray}
Therefore the following quantity is invariant under the $\lambda$-Lorentz 
transformations (\ref{eq6}):
\begin{eqnarray} &&
s^2=\left(\frac{v}{u}\right )^b\left (uv-y^2-z^2\right)\left(\frac{uv}
{uv-y^2-z^2}\right)^b=\nonumber\\ && v^{2b}(uv-y^2-z^2)^{1-b}=
\left(ct-x\right)^{2b}
\left(c^2t^2-x^2-y^2-z^2\right)^{1-b},
\label{eq10}
\end{eqnarray}
and can be considered as a generalization of the relativistic interval.
Accordingly, the space-time metric, invariant under the $\lambda$-Lorentz 
transformations (\ref{eq6}), has the form
\begin{equation}
ds^2=\left(c\,dt-dx\right)^{2b}
\left(c^2\,dt^2-dx^2-dy^2-dz^2\right)^{1-b}=\left(n_\nu dx^\nu\right)^{2b}
\left(dx_\mu dx^\mu\right)^{1-b},
\label{eq11}
\end{equation} 
where $n^\mu=(1,1,0,0)=(1,\vec{n}),\;\vec{n}^{\,2}=1$, is the fixed null-vector
that defines a preferred light-like direction in space-time.

As we see, in general, Einstein's two postulates do not lead to Lorentz 
transformations and the Minkowski metric. If the spatial isotropy is not 
assumed, they lead to more general $\lambda$-Lorentz transformations and 
Finslerian metric (\ref{eq11}) (applications of Finsler geometry in 
physics is considered, for example, in \cite{9,10,11}).

The obtained $\lambda$-Lorentz transformations (\ref{eq6}) correspond to the 
inertial reference frame $S^\prime$ which moves along the preferred direction 
$\vec{n}$.  If the inertial reference frame $S^\prime$ moves with the velocity 
$\vec{V}=c\,\vec{\beta}$ in an arbitrary direction, then the generalized 
Lorentz transformations that leave the Finsler metric (\ref{eq11}) invariant 
have the form \cite{12,13}
\begin{equation}
x^{\prime\mu}=D(\lambda)R^\mu_{\;\nu}(\vec{m};\alpha)\,L^\nu_{\;\sigma}
(\vec{V})x^\sigma,
\label{eq12}
\end{equation} 
where $L^\nu_{\;\sigma}(\vec{V})$ represents the usual Lorentz transformations,
$R^\mu_{\;\nu}(\vec{m};\alpha)$ is a rotation about the spatial direction
\begin{equation}
\vec{m}=\frac{\vec{n}\times\vec{\beta}}{|\vec{n}\times\vec{\beta}|}
\label{eq13}
\end{equation}
by the angle $\alpha$ such that
\begin{equation}
\cos{\alpha}=1-\frac{\gamma}{\gamma+1}\,
\frac{[\vec{n}\times\vec{\beta}]^2}{1-\vec{n}\cdot\vec{\beta}}\;.
\label{eq14}
\end{equation}
For the radius vector $\vec{r}$, the result of this additional rotation is 
given by the Euler-Rodrigues formula 
\cite{14}  
\begin{equation}
\vec{r}^{\,\prime}=\vec{r}+(\vec{m}\times\vec{r})\sin{\alpha}+
[\vec{m}\times(\vec{m}\times\vec{r})](1-\cos{\alpha}).
\label{eq15}
\end{equation}
Finally,  $D(\lambda)$ represents the dilatation
\begin{equation}
D(\lambda)x^{\mu}=
\lambda\,x^{\mu},
\label{eq16}
\end{equation}
with the scale-factor
\begin{equation}
\lambda=\left[\gamma(1-\vec{\beta}\cdot\vec{n})\right]^b.
\label{eq17}
\end{equation}
The explicit form of these generalized Lorentz transformations was obtained
in \cite{12,13}. They have the following form 
\begin{eqnarray} &&
x_0^\prime=\left[\gamma(1-\vec{\beta}\cdot\vec{n})\right]^b\gamma\,(x_0-
\vec{\beta}\cdot\vec{r}), \nonumber \\ &&
\vec{r}^{\,\prime}=\left[\gamma(1-\vec{\beta}\cdot\vec{n})\right]^b\left\{
\vec{r}-\frac{\vec{\beta}\,(x_0-\vec{n}\cdot\vec{r})}{1-\vec{\beta}\cdot
\vec{n}}-\right . \nonumber\\ && \left .
\vec{n}\left[\gamma\,\vec{\beta}\cdot\vec{r}+\frac{\gamma-1}{\gamma}
\,\frac{\vec{n}\cdot\vec{r}}{1-\vec{\beta}\cdot\vec{n}}+\frac{(\gamma-1)
\vec{\beta}\cdot\vec{n}-\gamma\beta^2}{1-\vec{\beta}\cdot\vec{n}}\,x_0
\right]\right\}.
\label{eq18}
\end{eqnarray}
If $\vec{\beta}$ and $\vec{n}$ are parallel, and the $x$-axis is along the 
velocity $\vec{\beta}$, transformations (\ref{eq18}) are reduced to the 
$\lambda$-Lorentz transformations (\ref{eq6}). 

The $\lambda$-Lorentz transformations (\ref{eq6}) were first derived by Lalan 
\cite{15,16,17}.  He also recognized that the metric, invariant with respect 
to the $\lambda$-Lorentz transformations, was of pseudo-Finslerian type. 
Generalized Lorentz transformations (\ref{eq18}) were first discovered by 
Alway \cite{18}. In his article, he cites Pars \cite{19} and mentions that the 
isotropic behavior of the clock, which was one of the Pars's assumptions, 
is generally not guaranteed. Alway's  transformations and the corresponding 
Finslerian space-time metric did not attract much attention, since it was 
immediately recognized \cite{20} that experimental observations severely 
constrain spatial anisotropy making the parameter $b$ practically 
zero. Some {\ae}ther-drift experiments imply $|b|<10^{-10}$,
while the Hughes-Drever type limits on the anisotropy of inertia can 
potentially lower the limit on the Finslerian parameter $b$ up to
$|b|<10^{-26}$, albeit in the model-dependent way \cite{21}. 

Generalized Lorentz transformations (\ref{eq18}) were soon rediscovered 
by Bogoslovsky \cite{12,22} who thoroughly investigated their physical 
consequences \cite{23,24,25}. A particular case of the
$\lambda$-Lorentz transformations (\ref{eq6}), corresponding to $b=1/2$,
was independently rediscovered by Brown \cite{26} and generalized to any
value of $b$ by Budden \cite{27}. Later they cite Bogoslovsky 
in \cite{28,29} and \cite{30}, but neither Lalan, nor Alway.  Such 
studies have largely remained outside the  mainstream, but recently they 
have gained more chances due to the advent of very special relativity.

\section{Very special relativity and Finsler geometry}
The Lorentz group does not have a 5-parameter subgroup. It has only one, up to 
isomorphism, 4-parameter subgroup, called SIM(2) (similitude group of the 
plane consisting of dilations, translations, and rotations of the plane) 
\cite{31,32}. If we add space-time translations to SIM(2), we get an 
8-parameter subgroup of the Poincar\'{e} group called ISIM(2). It is 
a semi-direct product of SIM(2) with the group of space-time translations.  
Cohen and Glashow proposed \cite{1} that the exact symmetry group of nature 
may not be the  Poincar\'{e} group, but its subgroup ISIM(2). 
The corresponding theory, very special relativity, breaks Lorentz symmetry 
in a very mild and minimal way. For amplitudes with appropriate analyticity 
properties, SIM(2) implies $CPT$ discrete symmetry, but it violates $P$ and 
$T$ discrete symmetries \cite{1}. However, in any theories of particle physics 
in which $P$, $T$ or $CP$ is conserved, Lorentz-violating effects of 
very special relativity will be absent, as the inclusion of any of these 
discrete symmetries extends the SIM(2) subgroup to the full Lorentz group 
\cite{1}. Since $CP$ violation is known to be small, Lorentz-violating effects 
in very special relativity are expected to be also small \cite{1}.

Generalized pure Lorentz boosts (\ref{eq18}), supplemented by rotations about 
preferred spatial direction $\vec{n}$ and by space-time translations, form
a 8-parameter group of isometries of the Finsler space-time with metric 
(\ref{eq11}) called DISIM$_b$(2) in \cite{21}. The very special relativity
symmetry group ISIM(2) is obtained from DISIM$_b$(2) by In\"{o}n\"{u} and 
Wigner contraction \cite{33,33A} in the $b\to 0$ limit. This limit is rather 
subtle. Indeed, although in the limit $b\to 0$ the Finsler metric (\ref{eq11}) 
is reduced to the Minkowski metric, the generalized Lorentz transformations 
(\ref{eq18}) are not reduced to the ordinary Lorentz transformations in this 
limit, because the preferred direction $\vec{n}$ can remain even in this limit.
In this case, the resulting symmetry group will not be the Lorentz group, 
but its 4-parameter subgroup SIM(2), consisting of  transformations 
(\ref{eq18}), with $b=0$, along with rotations about the preferred direction 
$\vec{n}$.

However, when $b=0$, that is when the space is isotropic, we have no reason 
to introduce the preferred light-like direction $n^\mu$ when deriving the
Lorentz transformations in the manner described above. Of course, we can 
calibrate the orientations of the spatial axes of inertial reference frames  
so that if in one of these reference frames the light beam has the 
direction $\vec{n}$, it will have the same direction in all inertial reference 
frames. In this case we will again arrive at the generalized Lorentz 
transformations (\ref{eq18}) (with $b=0$) instead of usual  Lorentz 
transformations. However, the resulting theory will still be the ordinary 
special relativity, but with the indicated special agreement on the 
orientations of the spatial axes \cite{12}. In this case we can choose 
light-lite vector $n^\mu$ arbitrarily and all such choices will be physically 
equivalent. In fact, for (\ref{eq18}) with $b=0$ to represent Lorentz symmetry,
rather than its SIM(2) subgroup, it is enough to require that the 
choices $\vec{n}$ and  $-\vec{n}$ are  equivalent. Indeed,  
$\vec{n}\to -\vec{n}$ corresponds to $\vec{m}\to -\vec{m}$ in (\ref{eq13}), 
and if we require the symmetry under rotations about $-\vec{m}$, 
the generalized Lorentz transformations (\ref{eq18}) with $b=0$ can be  
easily transformed into the ordinary Lorentz transformations by an additional 
rotation $x^{\prime\prime\,\mu}=R^\mu_{\;\nu}(-\vec{m},\alpha)x^{\prime\,\nu}$,
since $R^\mu_{\;\lambda}(-\vec{m},\alpha)R^\lambda_{\;\nu}(\vec{m},\alpha)=
\delta^\mu_\nu$.

In the light of the foregoing, we come to the conclusion that there is 
a natural possibility that fully respects the relativity principle, about how 
a very special relativity can arise, instead of the special relativity, in 
the description of reality. Namely, space-time can be Finslerian with the 
Lalan-Alway-Bogoslovsky metric (\ref{eq11}), but the parameter $b$ of this 
metric can be very small. The following analogy with the cosmological 
constant, described below, shows that this is indeed the most natural way
of introducing a preferred light-like direction into space-time theory. 

\section{Very special relativity in light of Segal's principle}
Although Minkowski, in his famous lecture ``Raum und Zeit'', never mentions 
Klein's Erlangen program of defining a geometry as theory of invariants of 
some group of transformations, a link between Minkowski's presentation of 
special relativity and Erlangen program was immediately recognized by Felix 
Klein himself \cite{35}. In 1954 Fantappi\'{e} rediscovered the connection 
and put forward a program which he himself called ``an Erlangen program for 
physics'': a classification of possible physical theories through their group 
of symmetries \cite{36,37}. 

In particular, Fantappi\'{e} discovered that the group of ``final relativity'' 
is not the Poincar\'{e} group, but the De Sitter group. The Poincar\'{e} group 
is just the limit of the De Sitter group when the radius of curvature of the 
De Sitter space-time turns to infinity, much like the Galilei group, which is 
the limit of the Poincar'{e} group, when the speed of light goes to infinity. 
In fact, these two examples are only part of a wider picture of possible 
kinematical groups and their interconnections within various limits, given 
later by H.~Bacry and J.~L\'{e}vy-Leblond \cite{38}. This picture can be
considered as a natural implementation of Fantappi\'{e}'s ``Erlangen program 
for physics'' and is based on the ideas of group contractions and 
deformations proposed by  In\"{o}n\"{u} and Wigner \cite{33} and by Irving 
Segal \cite{39}.  

Segal's principle  \cite{39,40,41} states that a true physical theory must 
be stable against small deformations of its underlying algebraic structure. 
For example, the Lie algebra of the inhomogeneous Galilei group is unstable 
in the sense of Segal, and its deformation leads to the Lie algebra of  the 
Poincar\'{e} group. As a result, the theory of relativity  based on the 
Poincar\'{e} group has a larger scope of validity than the theory of relativity
based on the Galilei group. However, the Poincar\'{e} Lie algebra is also 
unstable, and its small deformations lead to either de Sitter or anti-de 
Sitter Lie algebras \cite{38}. In light of this, it is not surprising
that it turned out that the asymptotic vacuum space-time is not Minkowski, but 
de Sitter space-time with non-zero cosmological constant. A really amazing
question is why the cosmological constant is so small. This is a profound 
question  of modern physics, and we still do not have a satisfactory answer 
to it.

As we have seen above, the very special relativity symmetry group ISIM(2) is 
not stable against small deformations of its structure, and a physically 
relevant deformation, DISIM$_b$(2), exists which leads to a Finslerian 
space-time. This was shown more formally in \cite{21}. In light of the Segal 
principle, we expect that, accordingly, the very special relativity cannot 
be a true symmetry of nature and should be replaced by DISIM$_b$(2). Then,
based on the analogy with the cosmological constant, it can be argued that 
if there really is a preferred light-like direction in nature, then the
Finslerian parameter $b$ will not be zero, but very small, so small that 
the corresponding Finslerian nature of space-time is unlikely to be detected 
in laboratory experiments.

\section{Conclusions}
In this note, we argued that if the very special relativity, and not the 
usual special relativity, is really implemented in nature, then most likely, 
space-time in the absence of gravity will have not Minkowski geometry, but 
Finsler geometry of the Lalan-Alway-Bogoslovsky type with the metric 
(\ref{eq11}). However, in this case the anisotropy parameter $b$ is expected 
to be very small, like the cosmological constant $\Lambda$. So small that 
it will be impossible to detect the effects of Finslerian nature of space-time 
in laboratory  experiments. Can then we repeat after Francesco Sizzi, the 
Florentine astronomer who opposed the discovery by Galileo of the moons of 
Jupiter, that such  anisotropy parameter ``can have no influence on the Earth, 
and therefore would be useless, and therefore do not exist''?\cite{42} 
Probably not, because the question of the true value of the anisotropy  
parameter $b$, like the cosmological constant problem, is of fundamental 
importance \cite{21}. 

\section*{Acknowledgments}
The work is supported by the Ministry of Education and Science of the Russian 
Federation.

\end{document}